\documentclass[10pt,aps,prd,reprint,nofootinbib,superscriptaddress]{revtex4-1}

\usepackage{graphicx}
\usepackage{amssymb}
\usepackage{amsmath}
\usepackage[colorlinks,breaklinks]{hyperref}
\hypersetup{colorlinks=true,allcolors=[rgb]{1,0.56,0}}

\newcommand{\orcid}[1]{\href{https://orcid.org/#1}{#1}}

\begin{document}

\title{Determining the Density of the Sun with Neutrinos}

\author{Peter B.~Denton}
\email{pdenton@bnl.gov}
\thanks{\orcid{0000-0002-5209-872X}}
\affiliation{High Energy Theory Group, Physics Department \\ Brookhaven National Laboratory, Upton, NY 11973, USA}

\author{Charles Gourley}
\email{charleshgourley@gmail.com}
\thanks{\orcid{0009-0008-3715-224X}}
\affiliation{High Energy Theory Group, Physics Department \\ Brookhaven National Laboratory, Upton, NY 11973, USA}
\affiliation{Rensselaer Polytechnic Institute, Troy, NY 12180-3590, USA}

\date{May 15, 2025}

\begin{abstract}
The discovery of solar neutrinos confirmed that the inner workings of the Sun generally match our theoretical understanding of the fusion process.
Solar neutrinos have also played a role in discovering that neutrinos have mass and that they oscillate.
We combine the latest solar neutrino data along with other oscillation data from reactors to determine the Sun's density profile.
We derive constraints given the current data and show the anticipated improvements with more reactor neutrino data from JUNO constraining the true oscillation parameters and more solar neutrino data from DUNE which should provide a crucial measurement of $hep$ neutrinos.
\end{abstract}

\maketitle

\section{Introduction}
There are two common means of determining the internal structure of the Sun.
The first class is known as standard solar models (SSM) which constrain standard physics within the Sun to observations of mass, luminosity, radius, age, and metallicity.
The second is helioseismology which takes seismic data from the surface of the Sun to model the dynamics of the interior, see e.g.~\cite{Gough:1996am,2016LRSP...13....2B}.
While both rely on modeling of the dynamics of the Sun to some extent, the latter is more model independent.
Unfortunately, it has limited sensitivity to the inner $\sim5-10\%$ of the Sun's radius.

The broad portfolio of neutrino oscillation experiments has provided key measurements of the fundamental physics parameters governing the neutrino sector, see e.g.~\cite{Esteban:2024eli,Capozzi:2021fjo,deSalas:2020pgw,Denton:2022een,Denton:2025jkt}.
By combining different classes of oscillation measurements of the same parameters, it is possible to use neutrino oscillations to probe environments that are otherwise inaccessible by easier to detect particles.
It has already been pointed out that measurements of neutrino oscillations or absorption with atmospheric, solar, supernova, and high energy astrophysical neutrinos can probe the structure of the Earth \cite{Nicolaidis:1990jm,Ohlsson:1999um,Jain:1999kp,Lindner:2002wm,Reynoso:2004dt,Akhmedov:2005yt,Akhmedov:2006hb,Winter:2006vg,Agarwalla:2012uj,Rott:2015kwa,Winter:2015zwx,Donini:2018tsg,DOlivo:2020ssf,Kumar:2021faw,Kelly:2021jfs,Denton:2021rgt,Capozzi:2021hkl,Upadhyay:2021kzf,DOlivoSaez:2022vdl,Maderer:2022toi,Upadhyay:2022jfd,Hajjar:2023knk,Raikwal:2023jkf,Petcov:2024icq,Upadhyay:2024gra,Jesus-Valls:2024tgd}.
We will show here how solar neutrinos can also probe into the physics happening inside the Sun and derive constraints on the density profile of Sun using neutrinos.
Photons tell us about the surface of the Sun today and the inner workings of the Sun hundreds of thousands of years ago after scattering many times through the hot Sun.
Neutrinos, on the other hand, tell us about the nature of the core of the Sun today and provide a relatively direct probe of its properties, subject to the physics of neutrino oscillations.

Using neutrinos to probe astrophysical environments is challenging mainly because neutrinos interact so weakly.
Moreover, neutrinos oscillate from production to detection so an independent measurement of the oscillation parameters is, in practice, required to extract information about the dense environment probed such as the inside of the Sun.

Detecting a statistically useful number of neutrinos requires very large detectors which typically come at a cost of higher thresholds and poor energy resolution.
Nonetheless, solar neutrinos have been measured across a range of energies with $\lesssim10\%$ precision \cite{Cleveland:1998nv,
Super-Kamiokande:1998qwk,Super-Kamiokande:1998zvz,Super-Kamiokande:1998oic,Super-Kamiokande:2001ljr,Super-Kamiokande:2001bfk,SNO:2001kpb,SNO:2002tuh,Super-Kamiokande:2002ujc,Super-Kamiokande:2005wtt,Super-Kamiokande:2008ecj,Super-Kamiokande:2010tar,SNO:2011hxd,Super-Kamiokande:2016yck,Super-Kamiokande:2023jbt}.
We will show how this data, along with future expected measurements, can be used to probe the density of the Sun.

This process of extracting the density of the Sun from solar neutrino data requires several interconnected components.
It leverages the fact that neutrino oscillations depend on the density of the Sun and that these oscillation parameters are independently measured on the Earth by KamLAND \cite{KamLAND:2013rgu}.
It also works because neutrinos from different nuclear processes in the Sun have different spectra and are produced in different regions of the Sun.
This allows for a measurement of the density profile of the Sun, an idea that seems to have first existed in 1997 in \cite{Balantekin:1997fr}.
Some time later, in 2013, it was pointed out that future improved solar neutrino measurements could potentially constrain the density of the Sun \cite{Lopes:2013sba}.
More recently, this concept has been investigated in the context of developing an algorithm to extra information about a supernova by demonstrating that consistent results can be achieved with a subset of the solar data \cite{Laber-Smith:2022eih,Laber-Smith:2024hbc}.

In section \ref{sec:review} we will provide the relevant background on solar neutrino oscillations and physics, as well as the role of the theory inputs.
We will discuss the data inputs in section \ref{sec:data} and present our results in section \ref{sec:results}.
We interpret our results in section \ref{sec:discussion} and conclude in section \ref{sec:conclusions}

\section{Solar Neutrino Review}
\label{sec:review}
In this section we review the physics of solar neutrinos, reactor neutrinos which measure the same oscillation parameters, and the theoretical prediction for the solar neutrino flux.
\subsection{Solar Neutrino Oscillations}
Solar neutrinos are produced in association with an electron.
Due to the dense environment in the Sun and the Wolfenstein matter effect \cite{Wolfenstein:1977ue}, the neutrinos produced as electron neutrinos do not propagate as they would in vacuum and, for high enough energies, asymptotically approach very close to a propagation eigenstate, $\nu_2$.

Specifically, the evolution of solar neutrinos is described by a Hamiltonian
\begin{equation}
H=\frac1{2E}\left[U
\begin{pmatrix}
0\\&\Delta m^2_{21}\\&&\Delta m^2_{31}
\end{pmatrix}U^\dagger+
\begin{pmatrix}
a\\&0\\&&0
\end{pmatrix}\right]\,,
\label{eq:H}
\end{equation}
where $E$ is the neutrino energy,
\begin{equation}
a=2\sqrt2G_FN_eE\,,
\end{equation}
is the strength of the matter potential, $G_F$ is Fermi's constant, $N_e$ is the electron number density at the production point, and $U$ is the PMNS matrix \cite{Pontecorvo:1957cp,Maki:1962mu} typically parameterized as
\begin{equation}
U=
\begin{pmatrix}
1\\
&c_{23}&s_{23}\\
&-s_{23}&c_{23}
\end{pmatrix}
\begin{pmatrix}
c_{13}&&s_{13}e^{-i\delta}\\
&1\\
-s_{13}e^{i\delta}&&c_{13}
\end{pmatrix}
\begin{pmatrix}
c_{12}&s_{12}\\
-s_{12}&c_{12}\\
&&1
\end{pmatrix}\,,
\label{eq:pmns}
\end{equation}
where we use the usual notation $s_{ij}=\sin\theta_{ij}$ and $c_{ij}=\cos\theta_{ij}$ and where the three mass eigenstates are defined as $|U_{e1}|^2>|U_{e2}|^2>|U_{e3}|^2$, see e.g.~\cite{Denton:2020exu,Denton:2021vtf}.
Then the solar neutrino probability is
\begin{equation}
P_{ee}^\odot=\sum_{i=1}^3|\hat U_{ei}|^2|U_{ei}|^2\,,
\label{eq:Pee}
\end{equation}
where $\hat U$ is the matrix that diagonalizes the Hamiltonian (eq.~\ref{eq:H}) at the production point of the neutrinos in the Sun.
Eq.~\ref{eq:Pee} leans on two key assumptions.
The first is that neutrinos decohere en route from the Sun to the Earth.
The distance from the Sun to the Earth is long enough for the wave packets to have separated and thus we have the incoherent sum of probabilities shown in eq.~\ref{eq:Pee}.
The second is that neutrinos transition adiabatically from the core of the Sun to its surface \cite{Mikheyev:1985zog}, a phenomenon known as the MSW (Mikheyev-Smirnov-Wolfenstein) effect.
The correction to this \cite{Parke:1986jy} is known to be very small for realistic neutrino parameters.

\begin{figure}
\centering
\includegraphics[width=\columnwidth]{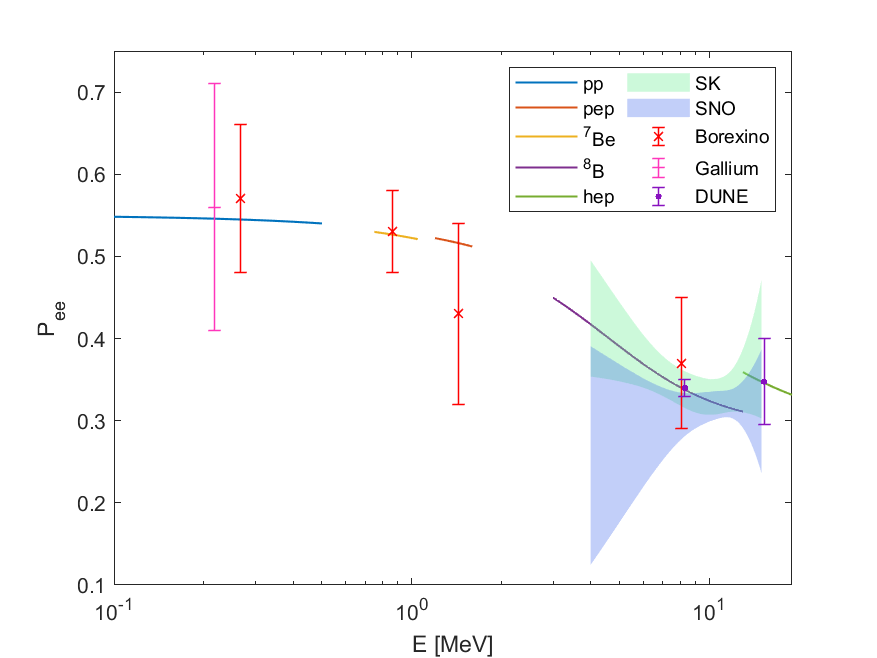}
\caption{Electron neutrino survival probability as a function of energy. The theory predictions are the solid lines plotted with cutoffs around their relevant flux dominance due to the different density associated with each neutrino source. The current experimental measurements are shown along with the expected improvements from DUNE.}
\label{fig:Pee}
\end{figure}

While the full solution to diagonalizing the Hamiltonian involves solving a cubic equation which has a notoriously complicated solution \cite{cardano}, we can get a fairly accurate picture of the physics in a two-flavor picture enhanced by some three-flavor corrections:
\begin{multline}
P_{ee}^\odot\approx\frac{c_{13}^4}2\left[1+\cos2\theta_{12}\vphantom{\frac{c_{13}^2\Delta m^2_{21}}{\sqrt{(\cos2\theta_{12}-c_{13}^2a/\Delta m^2_{21})^2+\sin^22\theta_{12}}}}\right.\\
\times\left.\frac{\cos2\theta_{12}-c_{13}^2a/\Delta m^2_{21}}{\sqrt{(\cos2\theta_{12}-c_{13}^2a/\Delta m^2_{21})^2+\sin^22\theta_{12}}}\right]+s_{13}^4\,.
\end{multline}
This resonant behavior was identified by Mikheyev and Smirnov \cite{Mikheyev:1985zog}.
In the low- and high-energy limits we have
\begin{equation}
P_{ee}^\odot\approx
\begin{cases}
1-\frac12\sin^22\theta_{12}\quad&E\ll1{\rm\ MeV}\\
\sin^2{\theta_{12}}&E\gg10{\rm MeV}
\end{cases}\,.
\end{equation}
The transition energy is given by the condition $a_{\rm res}\approx\cos2\theta_{12}\Delta m^2_{21}/c_{13}^2$.

We see that a measurement of the probability at low energies and high energies measures different functional forms of the same single parameter and thus one measurement predicts the other under the assumption of the standard three-flavor oscillation picture and a given SSM.
Thus solar neutrinos provide a valuable internal cross check of the consistency of the neutrino oscillation parameters.

\subsection{Long-baseline Reactor Neutrino Oscillations}
\label{sec:lbl reactor}
The same parameters that govern solar neutrino oscillations, $\theta_{12}$ and $\Delta m^2_{21}$, can also be probed in reactor experiments using a flux of $\bar\nu_e$ over long distances.
At leading order, such an experiment measures the probability
\begin{equation}
P_{ee}^{\rm reactor}\approx1-\sin^22\theta_{12}\sin^2\left(\frac{\Delta m^2_{21}L}{4E}\right)\,.
\end{equation}
Thus at a given baseline by measuring the depth and energy of the dip due to the first (highest energy) oscillation minimum, $\theta_{12}$ and $\Delta m^2_{21}$ can be determined in a straightforward manner.
We note, however, that long-baseline reactor experiments cannot determine the sign of $\Delta m^2_{21}$\footnote{Alternatively, one can say that long-baseline reactor neutrino experiments cannot determine the octant of $\theta_{12}$.
This choice depends on how one defines the mass eigenstates; see the text before eq.~\ref{eq:Pee}.}.
Determining this sign requires the matter effect which has a marginal impact on long-baseline reactor neutrinos \cite{Li:2016txk,Khan:2019doq}, but it plays an important role on solar neutrinos and is widely accepted to be positive.
In this paper we ignore any degenerate solutions involving the other sign on $\Delta m^2_{21}$.

\subsection{Standard Solar Model}
\begin{figure}
\centering
\includegraphics[width=\columnwidth]{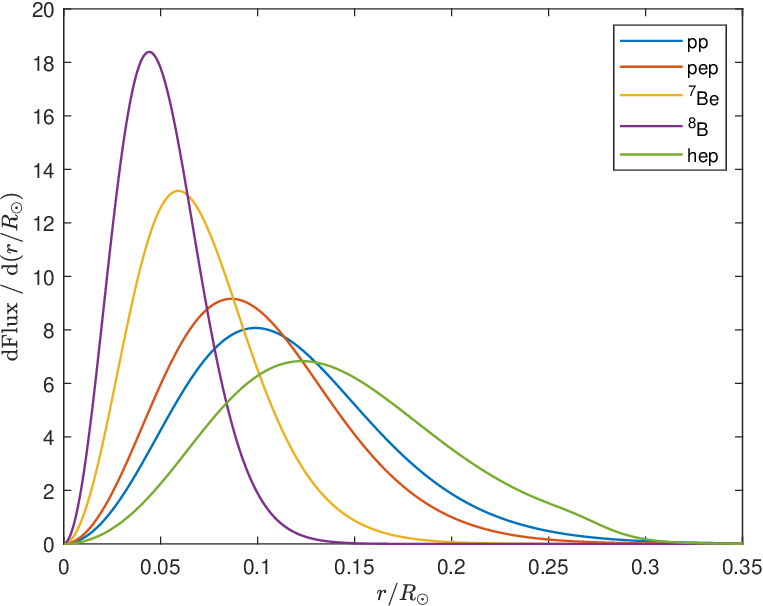}
\caption{The expected radial distribution of the production point of neutrinos from different processes, each of which are normalized to one, from \cite{Bahcall:2004pz}.}
\label{fig:radial distribution}
\end{figure}

The SSM is one method of modeling the Sun's interior based on theoretical calculations and simulations of the dynamics of a spherically symmetric ball of gas constrained by various electromagnetic observations.
Additionally, helioseismic data is used as well \cite{Turck-Chieze:2010rvs} and there is generally modest agreement among different approaches, although some minor tensions exist.
Helioseismic data has fairly precise (at the few percent level \cite{2025arXiv250307880K}) determinations of the density of the Sun except for the inner $r/R_\odot\lesssim0.05-0.1$ of the Sun \cite{2009ApJ...699.1403B,2016LRSP...13....2B} due to the lack of clear observations of gravity modes in the Sun \cite{2022A&A...658A..27B}, which is exactly the region where detected neutrinos are dominantly produced.
Relevant for neutrinos, the SSM predicts the energy spectra of neutrinos from various parts of the fusion chain and the radial distribution of where in the Sun these neutrinos are produced which are shown in fig.~\ref{fig:radial distribution}.
We take both of those as described by the theory.
The SSM also predicts the density profile of the Sun, expressed as the actual density as a function of radius along with the chemical composition, or as the electron number density, which is the quantity used in neutrino physics which is related to the other parameters inside the Sun via the temperature and pressure distributions.
It is this density profile that we treat as unpredicted and then extract from the neutrino data.

One could instead consider the variation of many details of the SSM including temperature, pressure, and composition in addition to density.
We focus on density alone for two reasons.
The first is because the data is and will be only good enough to provide information about one such distribution at a time.
The second is because the electron density, and thus by reasonable extrapolation the overall density, is what actually is probed by oscillations.
This makes our results more model independent in terms of solar modeling as well as new physics scenarios inside the Sun.

For the neutrino fluxes and radial profiles, we use that from \cite{Bahcall:2004pz}.
In \cite{Bahcall:2000nu} it was noted that a common, although somewhat outdated, simple means of estimating the electron density in the Sun is
\begin{equation}
\log_{10}\left(\frac{N_e\cdot{\rm\ cm}^3}{N_A}\right)=-4.58\frac r{R_\odot}+2.39\,,
\label{eq:Bahcall functional form}
\end{equation}
where $N_A$ is Avagadro's number and $R_\odot$ is the radius of the Sun.
This approximation is consistent with theoretical predictions for much of the Sun.
Although it is quite different from the full model prediction near the Sun's surface, neutrinos are not significantly produced from this region.
It also overestimates the density relative to the full prediction in the innermost region somewhat.
Nonetheless, it provides an important benchmark and a simple functional form to explore.

\section{Experimental Data}
\label{sec:data}
The complete set of existing solar neutrino measurements and expected future measurements presented in this section are shown in fig.~\ref{fig:Pee} along with the best fit probability for each separate neutrino flux calculated assuming the theoretically predicted density profile.
We note that, even though the density profile and oscillation parameters are taken to be equal across the entire plot, we see some nontrivial discontinuities, most notably between $^8$B and $hep$ probabilities.
This is due to the fact that $^8$B neutrinos come from the innermost part of the Sun where we expect the density is highest and the $hep$ neutrinos come from larger radii where we expect that the density is lower.
It is exactly this effect that we will leverage to determine the density profile of the Sun.
The Super-Kamiokande (SK) and SNO measurements of the $^8$B flux are shown as bands as described by those experiments.

\subsection{\texorpdfstring{$^8$B}{8B} data}
The most important data set for determining the electron density profile in the Sun comes from $^8$B data.
These neutrinos are high enough energy to be significantly modified by the presence of matter and they come from the innermost region of the Sun.
They are also the best measured part of the solar neutrino spectrum.

The most precise measurement comes from SK \cite{Super-Kamiokande:2023jbt} with an additional good measurement from SNO \cite{SNO:2011hxd}.
Their results are presented as a constraint on a parameterized functional form of the probability and covariance along with the overall $^8$B flux and the day-night measurement.
We do not include any information about the day-night effect in our analysis.
This is because nighttime neutrinos provide information about the density profile of the Earth.
Nighttime neutrinos do provide some additional information about $\Delta m^2_{21}$ and a little bit about $\theta_{12}$ but, given that KamLAND measures $\Delta m^2_{21}$ well, we do not include this information in our analysis.

We parameterize their results as shown in table \ref{tab:8B data} where the uncertainties reproduce the full uncertainties including both statistical and systematic uncertainties.
We have verified that this reproduces well the expected constraints on the standard oscillation parameters when the electron density profile predicted by the SSM is used.

We also use one data point from Borexino \cite{BOREXINO:2018ohr}
\begin{equation}
P^{^8\rm B}_{ee}=0.37\pm0.08\quad{\rm Borexino}\,,
\end{equation}
at an average energy $E=8.1$ MeV.

\begin{table}
\centering
\caption{Our extracted results from SK and SNO data.}
\label{tab:8B data}
\begin{tabular}{c|c||c|c}
\multicolumn{2}{c||}{SK} & \multicolumn{2}{c}{SNO}\\\hline
$E$ [MeV] & $P^{^8\rm B}_{ee}$ & $E$ [MeV] & $P^{^8\rm B}_{ee}$ \\\hline
7 & $0.354\pm 0.024$ & 7 & $0.296\pm 0.044$\\
9 & $0.332\pm 0.023$ & 9 & $0.312\pm 0.022$\\
11 & $0.331\pm 0.022$ & 11 & $0.320\pm 0.016$\\
13 & $0.349\pm 0.039$ & 13 & $0.320\pm 0.029$
\end{tabular}
\end{table}

\subsection{The intermediate lines: \texorpdfstring{$^7$Be}{7Be} and \texorpdfstring{$pep$}{pep}}
Two steps in the fusion process produce neutrinos with energies that are lines: $^7$Be at $E=0.862$ MeV and $pep$ at $E=1.44$ MeV.
Borexino has also measured these fluxes and finds
\begin{align}
P^{^7{\rm Be}}_{ee}&=0.53\pm0.05\quad{\rm Borexino}\,,\\
P^{pep}_{ee}&=0.43\pm0.11\quad{\rm Borexino}\,.
\end{align}
While the precision of these measurements is not that high, they do provide some information as the matter effect begins to play a small role in neutrino oscillations at $E\sim1$ MeV.

\subsection{\texorpdfstring{$pp$}{pp} data}
Detecting the low energy $E\lesssim0.5$ MeV $pp$ neutrinos is challenging.
Nonetheless, SAGE, GALLEX, and GNO have detected this flux by the use of gallium detectors \cite{GALLEX:1998kcz,SAGE:1999nng,GNO:2000avz,Gavrin:2001sz,SAGE:2009eeu}.
Borexino has also detected the $pp$ flux \cite{BOREXINO:2014pcl}.
All together, we use two data points for $pp$ neutrinos, one combined from the gallium experiments and the other from Borexino \cite{BOREXINO:2018ohr}:
\begin{equation}
P^{pp}_{ee}=
\begin{cases}
0.56\pm0.15\quad&{\rm gallium}\\
0.57\pm0.09&{\rm Borexino}
\end{cases}\,.
\end{equation}
We take the measurements as at $E=0.267$ MeV, the mean energy for Borexino's $pp$ measurement.
As the probability is expected to be quite flat in the region, we expect no significant impact due to the slightly different energies probed in each distinct kind of measurement of this flux.
The $pp$ data has little impact on constraining the density of the Sun directly, but does provide some information about $\theta_{12}$ which affects the higher energy solar neutrinos as well.

\subsection{Future DUNE solar \texorpdfstring{$hep$}{hep} data}
Future neutrino detectors are expected to improve measurements of solar neutrino parameters.
While Hyper-Kamiokande (HK) will have a significantly larger volume than SK, as well as improved PMTs, the smaller over burden weakens its sensitivity to solar neutrinos \cite{Hyper-Kamiokande:2018ofw,Barenboim:2023krl}.
The biggest improvement in solar neutrino data is expected to come from DUNE \cite{Capozzi:2018dat,DUNE:2020ypp,Meighen-Berger:2024xbx}.
We anticipate that HK and JUNO will also add useful information on solar neutrinos in the future and a combined analysis including all future experiments will be valuable to best determine the density profile of the Sun with neutrinos.
While a complete analysis from the experimental collaboration does not yet exist, we follow the analysis from \cite{Capozzi:2018dat} which should be approximately correct depending on the far detector details and total exposure.

DUNE is expected to be able to make an unprecedented measurement of $^8$B flux and the first measurement of the $hep$ flux.
Specifically, we assume that DUNE measures the probability expected by the best fit to the solar oscillation parameters and the SSM prediction for the electron neutrino density with the quoted uncertainties to find
\begin{align}
P^{^8\rm B}_{ee}&=0.340  \pm 0.010\qquad{\rm DUNE\ sensitivity}\\
P^{hep}_{ee}&=0.348\pm 0.052\qquad{\rm DUNE\ sensitivity}
\end{align}
where we assign energies $E=8.1$ MeV and $E=15$ MeV for the $^8$B and $hep$ fluxes, respectively.
The exact precision and effective energies will depend on a more sophisticated experimental treatment of the detector capabilities and backgrounds than currently exists, but we do not anticipate that it will significantly affect our sensitivity study.

\subsection{KamLAND and future JUNO data}
The approach described in section \ref{sec:lbl reactor} to measure the solar oscillation parameters using long-baseline reactor neutrinos was used by KamLAND \cite{KamLAND:2013rgu} and will be used by JUNO in the coming years \cite{JUNO:2022mxj}.
The current measurement from KamLAND provides a measurement of $\Delta m^2_{21}$ that is more precise than that extracted from all solar data, but the value of $\theta_{12}$ is less precise than that from solar data.

\begin{figure*}
\centering
\includegraphics[width=0.49\textwidth]{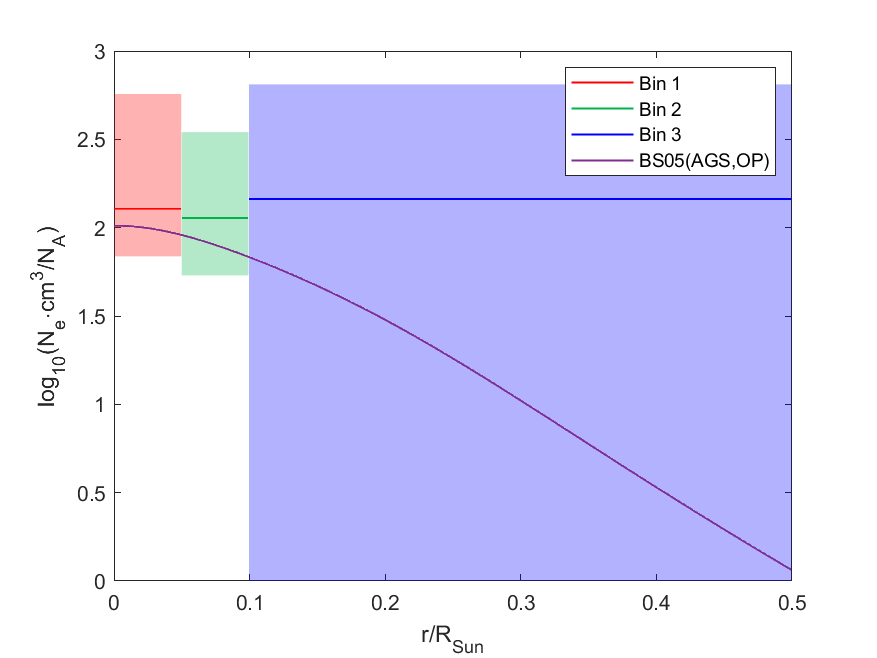}
\includegraphics[width=0.49\textwidth]{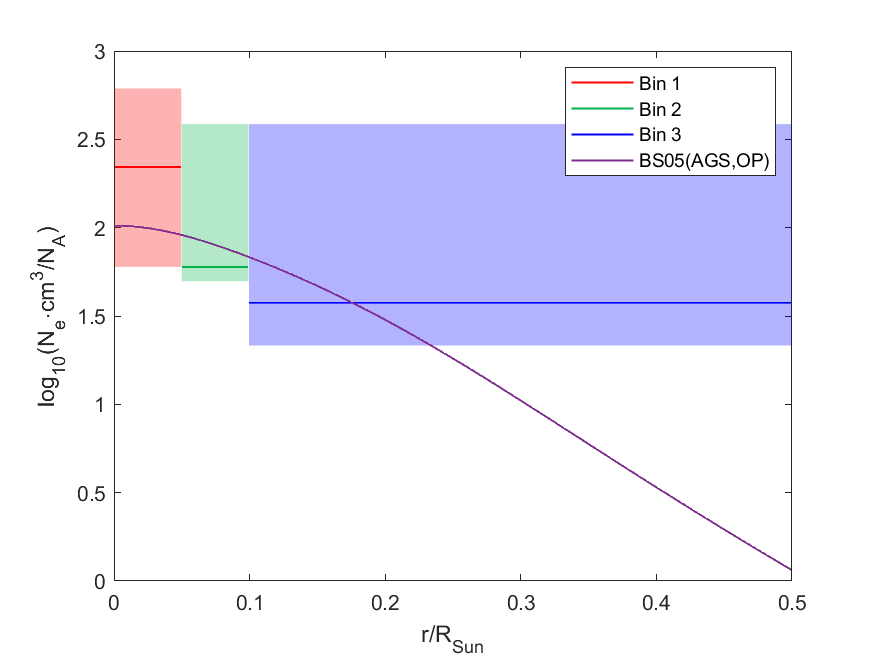}
\caption{Current measurements (left) and expected future constraints (right) of the Sun's density using the binned model along with the theory prediction from the SSM.
The shaded bands represent the $1\sigma$ uncertainties in each bin.}
\label{fig:bins 1D}
\end{figure*}

As JUNO aims to not only measure these same solar parameters with reactor neutrinos, but also measure the higher order frequencies at good precision, their precision on $\Delta m^2_{21}$ and $\theta_{12}$ will be well below the percent level and will easily be world leading both parameters in the coming years.
In fact, we have verified that it is not necessary to minimize the $\chi^2$ over these parameters as including the expected precision after six years of JUNO does not change our numerical results.

\vspace*{0.1in}

To summarize our discussion of the data, we will consider three classes of data sets.
\begin{itemize}
\item \textbf{Current}: This contains all relevant existing data, notably SK and SNO's $^8$B data, all Borexino solar neutrino data, and the various gallium experiments measuring $pp$ neutrinos.
This also contains KamLAND data which provides an independent constraint on the solar neutrino oscillation parameters.
\item \textbf{Current+JUNO}: The above data sets plus improved reactor measurements essentially fixing the solar oscillation parameters.
\item \textbf{Current+JUNO+DUNE}: The above data sets plus the expected improvements from DUNE's precise measurement of $^8$B neutrinos and their first measurement of $hep$ neutrinos.
\end{itemize}

\section{Modifications to the Sun's Density Profile}
The current understanding of the Sun's density profile comes from the SSM, a theoretical spherically symmetric model about the growth of stars fit to the current existing electromagnetic data.
We consider two means of quantifying the density profile in the Sun.

The first form of the density profile in the Sun that we consider is a binned profile of constant density in each of three bins.
The bins are defined as,
\begin{equation}
\frac r{R_\odot}\in
\begin{cases}
[0,0.05)&{\rm Bin\ 1}\\
[0.05,0.1)&{\rm Bin\ 2}\\
[0.1,0.5)\quad&{\rm Bin\ 3}
\end{cases}\,.
\end{equation}
There is no significant neutrino production for radii $\frac r{R_\odot}>0.5$.
The bin selections here are motivated by the fact that the $^8$B flux peaks at around $\frac r{R_\odot}\sim0.04-0.05$ while the $pep$ and $hep$ fluxes peak at higher radii.
Above $\frac r{R_\odot}\sim0.1$ the $^8$B flux is quite small and only the harder to measure fluxes are significantly contributing.

One potential issue with this binned approach is that, in principle, at each boundary the neutrino oscillation probability would experience a jump due to the sudden sharp transition.
We ignore this in our computations as the data does not support the existence of such a jump.
The density profile can be thought to be continuous by connecting sigmoid functions between the bins which would easily smooth away any deviations from adiabaticity.

The second means of quantifying the density profile in the Sun is a parametric fit in the form of eq.~\ref{eq:Bahcall functional form}.
We recall that the density at the origin and the derivative of the logarithm of the density are -4.58 and 2.39 in a simple fit to the SSM \cite{Bahcall:2000nu}, although this does slightly overestimate the density in innermost part of the Sun: $r/R_\odot\lesssim0.05$.
We use this functional form and fit the two parameters to the current neutrino data and expected future sensitivities.

\section{Results}
\label{sec:results}

\begin{figure*}
\centering
\includegraphics[width=\textwidth]{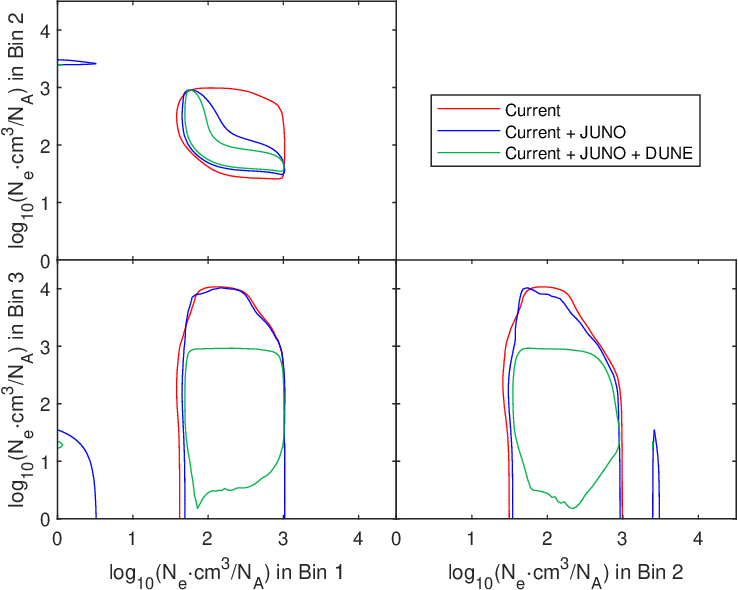}
\caption{The current measurements and expected future constraints in the binned model in the three different 2D projections at $\Delta\chi^2=6.18$ which is $2\sigma$ for 2 degrees of freedom assuming Wilks' theorem.}
\label{fig:bins 2D}
\end{figure*}

\begin{figure*}
\centering
\includegraphics[width=0.49\textwidth]{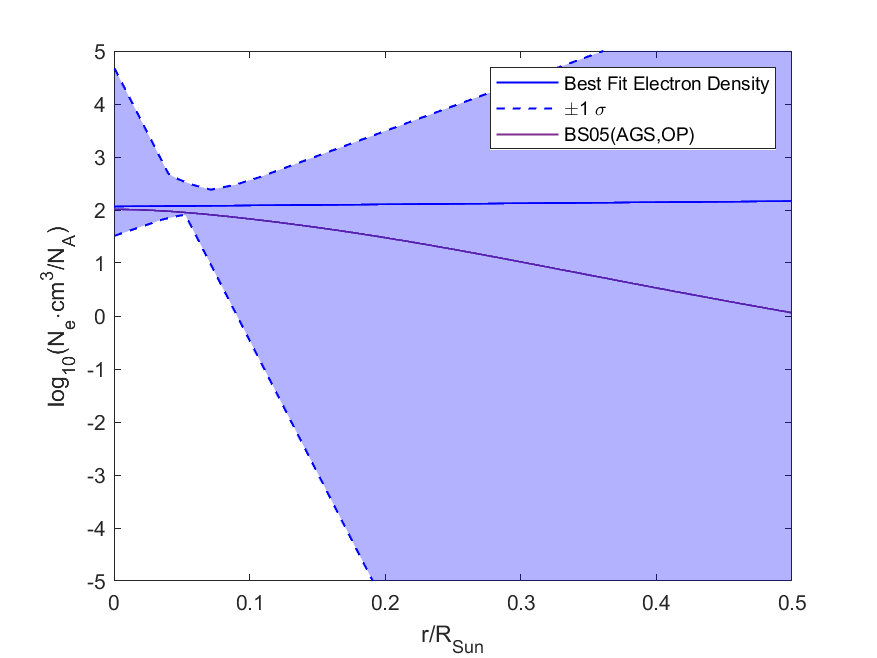}
\includegraphics[width=0.49\textwidth]{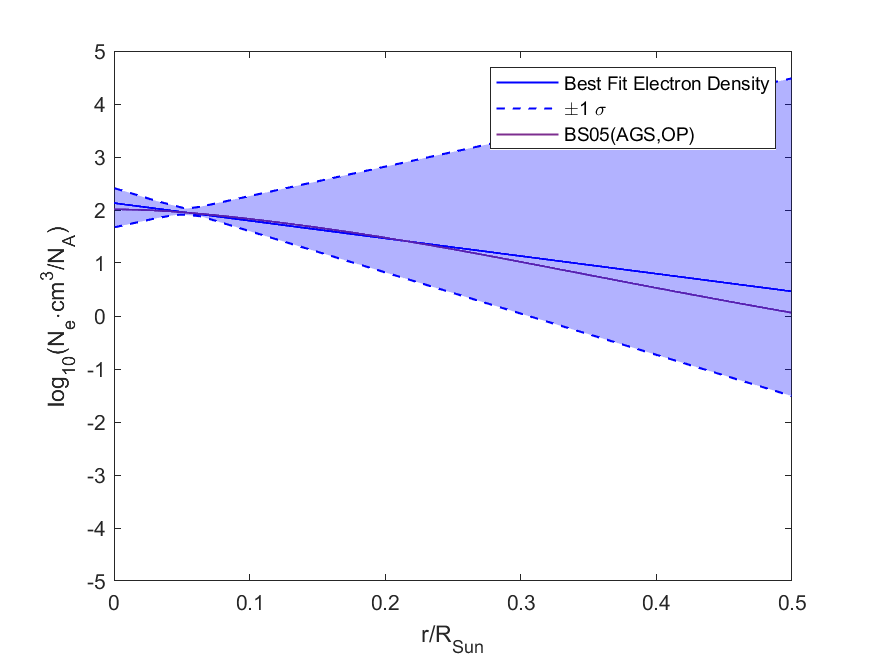}
\caption{Current measurements (left) and expected future constraints (right) of the Sun's density using the functional model along with the theory prediction from the SSM.}
\label{fig:functional 1D}
\end{figure*}

\begin{figure}
\centering
\includegraphics[width=\columnwidth]{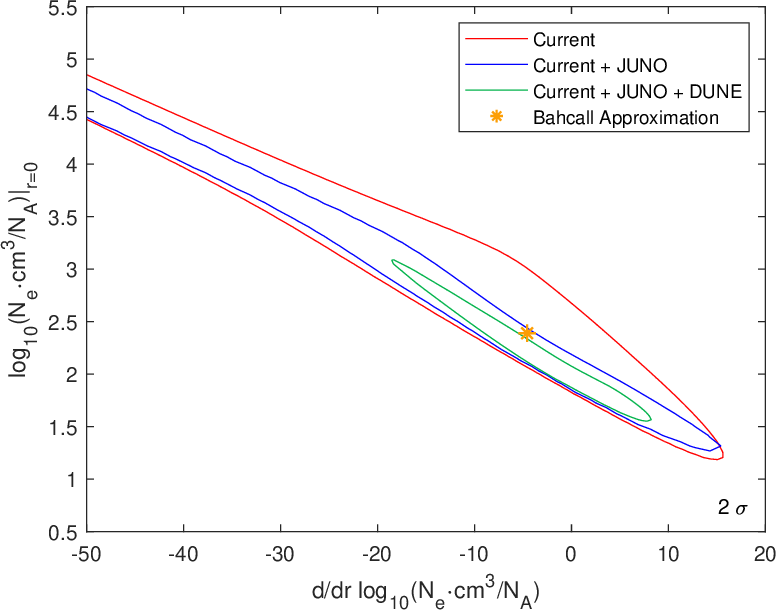}
\caption{Existing measurements and future sensitivities of the Sun's density using the functional model at $\Delta\chi^2=6.18$ which is $2\sigma$ for 2 degrees of freedom assuming Wilks' theorem.
The orange star indicates the theoretical approximation from eq.~\ref{eq:Bahcall functional form} from \cite{Bahcall:2000nu}.}
\label{fig:functional 2D}
\end{figure}

In order to constrain the density profile of the Sun, we construct a $\chi^2$ test statistic containing all the data described above for different models of the Sun's density.
We fix the non-solar oscillation parameters to best fit values \cite{Esteban:2024eli}.
The only non-solar parameter that plays any appreciable role in solar neutrino oscillations is $\theta_{13}$, but as that is measured extremely well \cite{DayaBay:2018yms,RENO:2018dro,DoubleChooz:2019qbj} by solar standards, it is not necessary to consider it as a nuisance parameter.
We have validated that our implementation of $^8$B data from SNO and SK reproduces the expected constraints on the standard solar parameters well when assuming the Sun's density profile is as expected.

First, we consider the binned form.
In this scenario, as discussed above, the density is constant across three bins: $r/R_\odot\in[0,0.05]$, $r/R_\odot\in[0.05,0.1]$, and $r/R_\odot\in[0.1,0.5]$.
There is no significant flux of neutrinos expected from larger radii.
Our results are shown in fig.~\ref{fig:bins 1D} across two different epochs: 1) current data including all solar data and KamLAND and 2) the addition of JUNO, which is expected to come online soon, and DUNE which is expected to come online after JUNO. It is worth noting that the improvements to current data from JUNO can not be seen in the 1-dimensional projections and is thus omitted in fig.~\ref{fig:bins 1D}.

We see that current data has already measured the density profile of the inner part of the Sun but in the larger radii $r/R_\odot\in[0.1,0.5]$ there is nearly no information on the density of the Sun from oscillations.
The addition of JUNO data will provide some improvement and the addition of DUNE solar data will significantly improve the picture due to the measurement of $hep$ neutrinos at high energy and at larger radii.

To better understand some of the more unusual aspects of the fit, we also show the preferred regions projected to two dimensions in fig.~\ref{fig:bins 2D} highlighting the non-trivial correlations among the different regions of the Sun.
We see that already with existing data there is a strong non-linear correlation between the preferred densities in the two inner regions.

Next we perform a fit to the functional form of the density profile of the Sun varying the slope in log space and the density at the center of the Sun.
Our results are shown in fig.~\ref{fig:functional 1D}.
We note that this parameterized prediction overestimates the density at small radii compared to the SSM, so it should be considered as a benchmark only.
We also show the preferred ranges of the parameters to the functional fit in fig.~\ref{fig:functional 2D} along with the simple prediction, where we notice that we find a closed region with future JUNO and DUNE data, but not before then.
We also see that the extremely unstable case of an increasing density profile will not be ruled out by neutrino data, something that can also be seen in fig.~\ref{fig:bins 1D}.

\section{Discussion}
\label{sec:discussion}
We see that while the improvements from additional detector information shown in fig.~\ref{fig:bins 1D}, seems to be modest in the binned density profile scenario, the 2D projection in fig.~\ref{fig:bins 2D} shows clearly that the new information does add additional information about the inner two bins disfavoring the scenarios where both bins are large.

The larger radii are only dominantly constrained with the addition of DUNE data, driven primarily from the $hep$ measurement.
The DUNE measurement significantly increases the lower limit on the density in the larger radius bin, but does not significantly change the upper limit from that without DUNE data.
This is because at densities larger than $\log_{10}(N_e\cdot{\rm\ cm}^3/N_A)\simeq3$ neutrinos experience the next higher resonance due to the atmospheric mass splitting.
Thus the upper limit is not significantly affected by the new DUNE $hep$ data but rather is ultimately constrained by the non-solar oscillation parameters and the fact that all the solar neutrino measurements are generally consistent with the regular single resonance picture.

We also see that there are degenerate solutions with the addition of the JUNO data. In reality, the degenerate solutions are always present for all data sets, but they are very close to the given $\chi^2$ threshold. Thus, the degenerate solution can be seen as a result of an increase in the minimum $\chi^2$ as the solar and reactor data slightly disagree, mainly in $\Delta m^2_{21}$. However, this increase in the minimum $\chi^2$ does not affect the degenerate solutions.

While we have chosen to fix some of the properties of the Sun to the SSM and let the electron number density vary, we could have asked different but related questions about the nature of the interior of the Sun.
For example, we could have instead assumed the theory prediction for the density profile, but considered different radial distributions of the production regions for each components of the Sun.
We could have also considered new physics scenarios that would affect the solar neutrino production in various non-trivial ways, see e.g.~\cite{Suliga:2020lir,Davoudiasl:2023uiq} for some examples.
Finally, we could also consider other exotic scenarios where dark matter could be captured in the Sun and contribute to the neutrino potential modifying solar neutrino propagation, possibly in an energy dependent way.

\section{Conclusions}
\label{sec:conclusions}
The interior of the Sun is not possible to directly probe with photons or charged particles.
The fusion chain that powers the Sun produces neutrinos of different energy spectra and at different radii inside the Sun.
This alone, however, is not enough to directly probe the density of the Sun with neutrinos.
Since neutrinos change flavor during propagation and that process depends on the local density through the MSW effect, by measuring the flux of $\nu_e$ at the Earth compared to the unoscillated predicted flux (which does not depend on the density of the Sun), it is possible to determine the strength of the matter potential where the neutrinos were produced and thus the density in that region.
This also requires, in practice, an independent measurement of the neutrino oscillation parameters which is provided by reactor neutrino experiments.

The constraints are driven by neutrino fluxes with energies near or above the transition region and those produced at different radii.
For this reason $^8$B neutrinos play an important role, as do the largely unmeasured $hep$ neutrinos.
Measuring each of those processes separately are important as they are both high enough energy to be significantly affected by the matter effect (unlike e.g.~$pp$ neutrinos) but come from quite different regions in the Sun.

Solar neutrino experiments have made good measurements of several different fluxes at the $\lesssim10\%$ level spanning $pp$, $pep$, $^7$Be, and $^8$B processes.
The oscillation parameters have been independently measured with reactor neutrinos by KamLAND.
In the future we can expect that JUNO will dramatically improve the reactor measurement to be essentially perfect on the level of precision of current and future measurements of solar neutrinos.
Beyond JUNO, we will also get significantly improved measurements of solar neutrinos from DUNE, notably an improved $^8$B measurement and likely the first highly significant $hep$ measurement.

We have shown that the existing data alone already provides some constraints on the density profile of the Sun, that it is consistent with the theoretical prediction of the SSM, and that future data from JUNO and DUNE will further improve this.

\section*{Acknowledgements}
We thank the anonymous referee for helpful comments.
PBD acknowledges support from the US Department of Energy under Grant Contract DE-SC0012704.
This project was supported in part by the U.S.~Department of Energy, Office of Science, Office of Workforce Development for Teachers and Scientists (WDTS) under the Science Undergraduate Laboratory Internships Program (SULI). 

\bibliography{main}

\end{document}